\documentclass{ws-procs9x6}
\usepackage{epsfig}
\def\be{\begin{equation}}
\def\ee{\end{equation}}
\def\bea{\begin{eqnarray}}
\def\eea{\end{eqnarray}}

\def\del{\partial}

\def\fr{\frac}

\begin{document}

\title{Gribov's Light Quark Confinement Scenario}

\author{C.\ EWERZ}

\address{Institut f\"ur Theoretische Physik, 
Universit\"at Heidelberg\\ 
Philosophenweg 16, 69120 Heidelberg, 
Germany\\
E-mail: c.ewerz@thphys.uni-heidelberg.de}


\maketitle

\abstracts{I give a brief description of Gribov's light quark 
confinement scenario and of the Gribov--Dyson--Schwinger 
equation for light quarks. It is shown that the Green function 
obtained from this equation exhibits chiral symmetry breaking. 
}

\section{Gribov's Confinement Scenario}

An interesting physical mechanism for explaining the 
phenomenon of confinement was proposed by V.\,N.\ 
Gribov in Refs.\ \refcite{Gribov:1998kb,Gribov:1999ui}. 
This mechanism is based on the idea the colour charges 
in QCD are  supercritical. The analogous phenomenon is 
well known in QED where a pointlike charge $Z>137$ 
is unstable and captures an electron from the vacuum 
while a positron is emitted. In QCD a similar phenomenon 
can occur due to the existence of very light quarks. If 
already a single quark has a supercritical colour charge 
this leads to a dramatic change in the vacuum structure 
of light quarks (see Refs.\ \refcite{Gribov:1998kb,Ewerz:2000qb}). 
The properties of the new vacuum structure are such that 
a single quark is unstable and can be observed as a resonance but 
not as an asymptotic state. Interestingly, confinement arises in 
this picture due to the Pauli principle rather than due to 
very strong forces. 

\section{The Gribov-Dyson-Schwinger Equation for Light Quarks}

A new approximation scheme for the treatment of the exact 
Dyson--Schwinger equation for quarks was suggested in Ref.\ 
\refcite{Gribov:1998kb}. This scheme is motivated by the physical 
picture of supercritical charges, but is also interesting in its own right. 
One uses Feynman gauge in which the Green function of the 
gluon has the form $D_{\mu\nu}(k) = - \alpha_s(k)g_{\mu\nu}/k^2$,  
and one assumes the effective strong coupling constant $\alpha_s(k)$ 
to be a slowly varying function that remains finite at low momenta --- 
an assumption that is in agreement with recent results obtained 
in the dispersive approach to power corrections\cite{Dokshitzer:1995qm}. 
The choice of the Feynman gauge allows one to collect in a very 
elegant way those contributions to the Dyson--Schwinger equation 
which are most singular in the infrared. Clearly, it is this 
region in which the dynamics of chiral symmetry breaking and 
confinement mainly resides. The corresponding approximation scheme 
then leads to the Gribov--Dyson--Schwinger equation for the 
Green function $G$ of a light quark, 
\be
\frac{\del^2}{\del q^\mu \del q_\mu}  \,
G^{-1}(q) = C_F \fr{\alpha_s(q)}{\pi} 
\,(\del^\mu G^{-1}(q)) \,G(q) \,(\del_\mu G^{-1}(q)) 
\,,
\label{gds}
\ee
where $C_F=4/3$. A very interesting observation is that 
the running coupling $\alpha_s(q)$ in this equation ensures 
that at large space--like momenta 
the solutions exactly reproduce 
the correct mass and wave function renormalization 
as it is known in perturbation theory. 
In principle one can also include subleading terms in the equation, 
i.\,e.\ terms which are less singular in the infrared. 

\section{Chiral Symmetry Breaking and Critical Coupling}

\begin{figure}
\begin{center}
\input{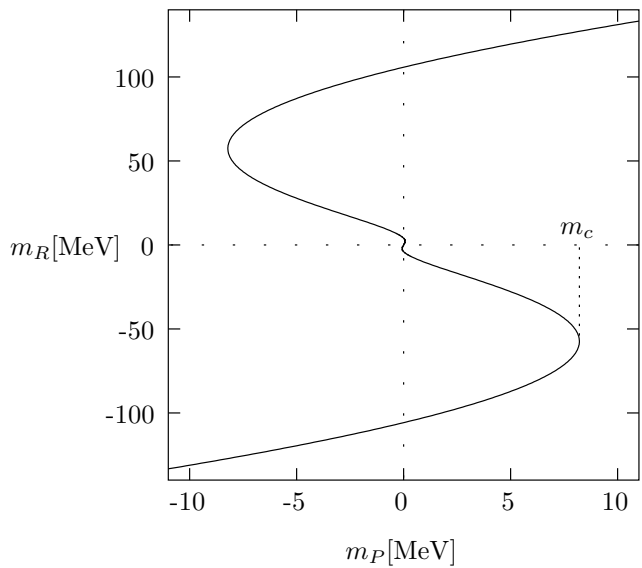}
\caption{\label{fig:m2m}
Relation between perturbative mass $m_P$ and renormalized 
mass $m_R$ 
(obtained with a model for $\alpha_s(q)$ motivated by Ref.\ 
\protect\refcite{Dokshitzer:1995qm})
}
\end{center}
\end{figure}

A detailed study of the Gribov--Dyson--Schwinger equation 
was performed in Ref.\ \refcite{Ewerz:2000qb}. It turns out that 
the resulting Green function shows a characteristically 
different behaviour if the effective strong coupling constant 
$\alpha_s(q)$ exceeds a critical value in some interval in the 
small momentum region. The critical coupling $\alpha_c$ 
can be determined analytically by considering asymptotic 
expansions of eq.\ (\ref{gds}) and is 
\be
\label{critcoupl}
\alpha_c = \frac{\pi}{C_F} \left( 1- \sqrt{\fr{2}{3}}\,\right)
\simeq 0.43
\,.
\ee
The change in the behaviour of the Green function corresponds 
to chiral symmetry breaking. It is best seen in the behaviour of 
the dynamical mass function $M(q^2)$ of the quark at 
space--like momenta, $q^2<0$. Let us define 
a `perturbative' mass $m_P=M(\lambda^2)$ at some large 
momentum scale $\lambda$ where perturbation theory holds. 
Further we define a `renormalized' mass $m_R=M(0)$ as the 
low--momentum limit of the dynamical mass function. 
If the coupling $\alpha_s$ remains subcritical in the infrared 
the relation between $m_P$ and $m_R$ is monotonic, 
and $m_R$ vanishes for vanishing $m_P$. If the coupling 
becomes supercritical in the infrared, $\alpha> \alpha_c$, 
that behaviour changes dramatically. The relation between 
$m_P$ and $m_R$ for this case is shown in fig.\ \ref{fig:m2m}. 
One sees that the renormalized mass remains finite 
even for vanishing perturbative mass, clearly signalling 
 chiral symmetry breaking. 

\section{Outlook}

One expects that the confinement of quarks is reflected 
in the analytic structure of their Green function. 
The analytic structure resulting from the 
Gribov--Dyson--Schwinger equation (\ref{gds}) has been 
shown\cite{Ewerz:2000qb} to exhibit poles and cuts 
at positive $q^2$, and hence does not correspond 
to confined quarks. This result had been 
anticipated\cite{Gribov:1998kb} because this equation does not 
take into account the back-reaction of the pions 
which are generated as Goldstone bosons in the process 
of chiral symmetry breaking. This back-reaction is 
expected to be crucial for the emergence of confinement, 
and there are indications that its inclusion in the 
equation in fact induces an analytic behaviour of the Green 
function corresponding to confined quarks\cite{Gribov:1999ui}. 


\end{document}